\documentclass[12pt]{article} 

\usepackage{graphicx}
\usepackage{amsmath}
\usepackage{amssymb}
\usepackage{hyperref}
\usepackage[height=8.8in,width=6.45in]{geometry}
\usepackage{tikz}
\usepackage{cite}
\usepackage{amscd}
\usepackage{setspace}
\usepackage{bbm}
\usepackage{dsfont}
\usepackage{cancel}
\usepackage{relsize}
\usepackage[utf8]{inputenc}
\usepackage{xcolor}

\numberwithin{equation}{section}

\newcommand{\e}{\mathrm{e}}

\begin{document}
\begin{titlepage}

\begin{flushright}

\end{flushright}

\vskip 3cm

\begin{center}
{\Large \bf
Bosonic quantum integrable systems and gauge theories}

\vskip 2.0cm

 Wei Gu  \\

\bigskip
\begin{tabular}{cc}
 Max-Planck-Institut f\"ur Mathematik\\
  Vivatsgasse 7\\
   D-53111 Bonn, Germany
 \end{tabular}

\vskip 1cm

\textbf{Abstract}
\end{center}








In this paper we apply two-dimensional supersymmetric gauge theories to directly construct a new Bethe ansatz for the wavefunctions of the q-boson hopping model, and 
then derive the q-boson algebras from this ansatz.

\medskip
\noindent

\bigskip
\vfill
\end{titlepage}

\setcounter{tocdepth}{2}
\tableofcontents
\section{Introduction}

Integrable systems have a long history in gauge theories, see for example
\cite{Nekrasov:2009rc, Nekrasov:2009uh, Nekrasov:2009ui,Beisert:2010jr,Costello:2017dso,Costello:2018txb, Ashwinkumar:2018tmm, Ashwinkumar:2019mtj}.
It was observed by Nekrasov and Shatashvili in \cite{Nekrasov:2009rc, Nekrasov:2009uh, Nekrasov:2009ui} that the Bethe ansatz equations of Yang-Baxter integrable systems are the vacuum equations of an associated 2d ${\cal N}$=(2,2) gauge theories. See also their earlier work \cite{Moore:1997dj} with Moore for an initial observation of this correspondence, which was further clarified by Gerasimov and Shatashvili in \cite{ Gerasimov:2006zt, Gerasimov:2007ap}. Certainly, their dictionary contains more information centered around the Yang-Baxter integrable system. Furthermore, some new progress has been made recently. See, for instance, \cite{Bullimore:2017lwu, Dedushenko:2021mds, Bullimore:2021rnr}. 


This paper is a followup to the author's work \cite{Gu:2022ugf} discussing how Heisenberg spin chains can emerge from an associated 2d ${\cal N}$=(2,2) gauge theories under renormalization group flow.
Most of the models discussed in \cite{Gu:2022ugf} are spin chains with local fermionic degrees of freedom. However, there exist many bosonic integrable systems studied in the literature such as the phase model studied, for example, in \cite{Bogolubov:1996} (and references therein), its generalization to the q-boson hopping model \cite{Bogoliubov:1997soj}, and others \cite{Faddeev:1996iy, Beisert:2008tw}\footnote{We apologize that we could not cite all the related papers due to the limitation of space.}. So it would be interesting to know which gauge theories correspond to them. 
The gauge theory for the q-boson model was studied by Okuda and Yoshida in \cite{Okuda:2012nx, Okuda:2013fea} following the results discovered by Korff and Stroppel \cite{Korff:2010} and Korff \cite{Korff:2011}. They reproduced the Bethe ansatz equations of the q-boson model in \cite{Bogoliubov:1997soj,Korff:2011} from the vacuum equations of the $G/G$ gauged WZW-(matter) model. Finally, we comment that in order to explain the hopping terms of the q-boson model following the framework discussed in \cite{Gu:2022ugf}, one needs a ``big theory" for the $G/G$ gauged WZW-(matter) model. 
This theory is a three-dimensional theory, the \textit{q-deformed Chern-Simons matter theory}, which was initially investigated by Gukov and Pei in \cite{Gukov:2015sna}. It reduces via Kaluza-Klein on the spacetime  $\Sigma \times S^1$ to the $G/G$ gauged WZW-matter model after integrating out fundamental matter fields.

In this paper, we continue to study the correspondence between the q-deformed boson model and the associated gauge theory. The q-deformed boson model is a lattice theory with $n$ sites, each with a creation operator $Q^{\dagger}_{i}$ and an annihilation operator $Q_i$ on the $i$-th site.  In the gauge theory correspondence, these $n$ sites correspond to critical loci of the effective superpotential, and the sites form a module under a global $\mathbb{Z}_{n}$ symmetry of the gauge theory. Furthermore, the Hamiltonian and the commutator of operators can be read off from the Bethe ansatz as discussed in section \ref{QDA}. In this paper, 
we find a new Bethe ansatz for the Bethe eigenvectors of the q-deformed boson model inspired by gauge theories. The existing literature for this model has focused on the \textit{algebraic Bethe ansatz}. Our new Bethe ansatz
\begin{equation}\label{qbwf1}
   \psi\left(p_{1},\cdots, p_{k}\right) = A_{k}\sum_{1\leq j_{1}\leq\cdots \leq j_{k}\leq n}\sum_{{\cal W}\in S_{k}}S_{{\cal W} \left(a\right)}e^{\sum^{k}_{a=1}ip_{{\cal W} \left(a\right)}j_{a}}|j_{1}\leq\cdots\leq j_{k}\rangle_{k}
\end{equation}
is very similar to the \textit{coordinate Bethe ansatz} discussed in the spin-$\frac{1}{2}$ chain system except that in our case we are allowed to put different bosons on the same site. This results in analogues of contact terms which do not merely describe the UV physics of the eigenstates, as they are relevant for IR physics. Then using this wave function, we find a q-deformed algebra emerging from the consistency requirement of the integrable system, namely, that the wave function be an eigenstate of the Hamiltonians. This new discovery encourages us that gauge theory can teach us something more about the integrable system. A one-form global symmetry (the center symmetry of the dual theory) and the Weyl symmetry of $G$ in the $G/G$ gauged WZW-(matter) model play a crucial role in our observation \cite{Witten:1993xi}. 
We will then read off the symmetries from the vacuum equations. 

In section \ref{PGT}, we briefly review some basics of $G/G$ gauged WZW-(matter) theory and the q-deformed Chern-Simons. We will give a new derivation for the partition function of $SU(n)_{k}$ $G/G$ gauged WZW theory by using the level/rank duality \cite{Naculich:1990pa, Hsin:2016blu} between 3d Chern-Simons theories and their KK-reductions are the gauged WZW theories \cite{Blau:1993tv}  
\begin{equation}\label{lrd}
  SU(n)_{k}\leftrightarrow U(k)_{-n,-n}.
\end{equation}
To match the numerically conjectured partition function of $SU(n)_{k}$ $G/G$ gauged WZW-matter theory proposed in \cite{Okuda:2013fea}, we expect there should be a Seiberg-like duality between $G/G$ gauged WZW-matter theories as follows:
\begin{equation}\label{lrd2}
  SU(n)_{k}+?\leftrightarrow U(k)_{-n,-n}+ \textrm{adjoint matter}.
\end{equation}

In section \ref{NBAQ}, we will give physics reasons for finding the new Bethe ansatz. It mainly follows the construction of the wave function in spin chain discussed in \cite{Gu:2022ugf}. In section \ref{QDA}, we use the proposed Bethe ansatz to derive the q-deformed boson algebra following the usual consistency conditions in the integrable system. Some identities are needed in this derivation that we put in Appendix \ref{IDAP}.

\section{Some preliminaries}\label{PGT}
In this section, we briefly introduce the $G/G$ gauged WZW-(matter) models \cite{Okuda:2012nx, Okuda:2013fea} and the q-deformed complex Chern-Simons \cite{Gukov:2015sna}. We mainly focus on the aspects that we will use in this paper.

\subsection{$G/G$ gauged WZW models}\label{GWZW}
The $G/G$ gauged WZW model is a special type of the gauged WZW model. Let us denote the 2d spacetime to be $\Sigma_{h}$, which is a genus $h$ Riemann surface; the complex structure determines the Hodge duality operators $\ast$ on one-forms.  The field variables include the group-valued defined via the map g: $\Sigma_{h}\rightarrow G$ and the gauge connection $A$. The action of the $G/G$ gauged WZW model is
\begin{eqnarray}\label{GWZWA}
  S_{G/G} &=& -\frac{1}{8\pi} \int_{\Sigma_{h}}{\rm tr}g^{-1}d_{A}g\ast g^{-1}d_{A}g-i\Gamma(g,A), \\\nonumber
  \Gamma(g,A) &=&  \frac{1}{12\pi}\int_{N}{\rm tr}\left(g^{-1}dg\right)^{3}-\frac{1}{4\pi}\int_{\Sigma_{h}}{\rm tr}\left(A\left(dgg^{-1}+g^{-1}dg\right)+Ag^{-1}Ag\right),
\end{eqnarray}
where $N$ is a three-manifold with $\partial N=\Sigma_{h}$. The partition function of the level $k$ $G/G$ gauged WZW model has a manageable form derived in \cite{Blau:1993tv} via the abelianization:
 \begin{equation}\label{GWZWPF}
   Z\left(S_{G/G}(\Sigma_{h},k)\right)=\int D[\phi, A]\exp(i(k+\rho)S_{\phi F}(\phi, A)),
 \end{equation}
where $\rho$ is the dual Coxeter number of $G$, $\phi$ is a compact scalar field taking values in the maximal torus $T$ of $G$, and $A$ is the reduced gauge field valued in the Lie algebra \textbf{t} of $T$.  The compact abelian BF action and the measure are
\begin{equation}\label{BFA}
  S_{\phi F}=\frac{1}{2\pi}\int_{\Sigma_{h}}{\rm tr}\phi F_{A}
\end{equation}
and
\begin{equation}\label{Mea}
  D[\phi, A]=D\phi DF_{A}\det\left(\textbf{1}-{\rm Ad}(g)\right)^{1-h}
\end{equation}
respectively. We mainly focus on the sphere partition function in this paper. Now, let us apply the above formula to several concrete examples.\\

\noindent $\underline{U(1)_{N}}$:\\
The partition function is 
\begin{equation}\label{}
  Z(S_{U(1)_{N}}(\Sigma_{h}))=\sum^{+\infty}_{n=-\infty}\int^{2\pi}_{0}\frac{d\phi}{2\pi}\delta\left(N\frac{\phi}{2\pi}-n\right).
\end{equation}
Since the domain of integration is $[0,2\pi]$, only a finite number of values of $n$ contribute to the partition function, namely
\begin{equation*}
  0\leq n\leq N-1.
\end{equation*}
We can express these contributions in terms of the original group variable $g=e^{i\phi}$ as solutions obeying
\begin{equation}   \label{eq:vac-eqn}
  g^{N}=1,
\end{equation}
which can be interpreted as the vacua of the theory.
So the \textit{vacuum equation}~(\ref{eq:vac-eqn}) indeed has $N$ solutions, which corresponds to the number of primary fields of the $U(1)_{N}$ GWZW model. Finally, we comment that the above ``vacuum equation" manifests the $\mathbb{Z}_{N}$ symmetry of the $U(1)_{N}$ GWZW model. \\

\noindent $\underline{U(k)_{N-k, N}}$:\\
This case was studied in \cite{Witten:1993xi}, which showed the correspondence between the Verlinde algebra and the quantum cohomology of the Grassmannian. After the abelianization, the nonabelian level $N-k$ of $SU(k)$-part will be shifted to $N$. Then the partition function can be easily derived:
\begin{eqnarray}\label{NAPF}
   Z(S_{U(k)_{N-k, N}}(\Sigma_{h})) = \frac{1}{k!}\sum^{\infty}_{n_{1},\cdots, n_{k}=-\infty}&&\int\prod^{k}_{a=1}\frac{d\phi_{a}}{2\pi}\prod^{k}_{a,b=1;
   a\neq b}\left(1-e^{ i(\phi_{a}-\phi_{b})}\right)^{1-h}\times \\\nonumber
   && \prod^{k}_{a=1}\delta\left(\frac{N}{2\pi}\phi_{a}-n_{a}+\frac{k-1}{2}\right) .
\end{eqnarray}
The delta functions in the partition function above are the same as those in the partition function of the $U(1)^{k}_{N,\cdots,N}$ gauged WZW model. They express a localization on the solutions of the \textit{vacuum equations} of the theory, which can be expressed in terms of the group variables $g_{a}=\exp( i \phi_{a})$:
\begin{equation}\label{GRVE}
  \left(g_{a}\right)^{N}=(-1)^{k-1},\quad\quad {\rm for} \quad a=1,\cdots,k.
\end{equation}
The measure factor in Equ'n (\ref{NAPF})
\begin{equation}\label{VADM}
  \prod^{k}_{a,b=1;
   a\neq b}\left(1-g_{a}\left(g_{b}\right)^{-1}\right)
\end{equation}
  must be non-vanishing as it comes from the Jacobian of the change of variables from the nonabelian group variable to its maximal torus part. So one must impose the following constraint
\begin{equation}\label{ECL}
  g_{a}\neq g_{b}, \quad\quad {\rm if}\quad a\neq b
\end{equation}
on the reduced group variables. Finally, we expect that the nonabelian enhanced locus will not contribute to the partition function \cite{Blau:1993tv, Gu:2018fpm}.\\

\noindent $\underline{SU(n)_{k}}$:\\
This is one of the models we are primarily interested in this paper. We will show that it corresponds to the phase model \cite{Okuda:2012nx, Korff:2010}. The partition function was initially computed in \cite{Blau:1993tv}, and they found only the following field configurations
\begin{equation*}
  \phi_{a}=\frac{2\pi n_{a}}{k+n},\quad\quad n_{a}>0,\quad \sum_{a}n_{a}<k+n.
\end{equation*}
contribute to the partition function. The $\mathbb{Z}_{n}$ center symmetry of the $SU(n)$ group is not manifest in this presentation. However, it is manifest in the dual theory. It was initially studied in \cite{Okuda:2012nx} by matching the statements proved in \cite{Korff:2010}. The partition function was derived in \cite{Okuda:2012nx}, and we will give a different derivation here.

The level/rank duality  we will use in the computation is
\begin{equation}\label{LRD}
  SU(n)_{k}\leftrightarrow U(k)_{-n,-n}.
\end{equation}
The negative sign of the level on the right-hand side can be understood from the Gauge/Bethe correspondence as follows. Recall that this Seiberg-like duality maps to the (parity)-symmetry in the integrable system \cite{Gu:2022ugf}\footnote{In an earlier paper \cite{Gaiotto:2013bwa} (see also \cite{Benini:2014mia}), Seiberg duality was treated as the ``Weyl-reflection symmetry" in the associated integrable system.}, then use the fact that the parity transformation  changes the sign of the level.

We can, in fact, obtain the partition function by using the one of $U(k)_{n, (n+k)}$ gauged WZW model computed in (\ref{NAPF}). One can easily observe that the difference between the two theories is a $\mathbb{Z}_{k}$ orbifolded $\left(\frac{U(1)_{k^{2}}}{\mathbb{Z}_{k}}\right)$ gauged WZW model with the Lagrangian
\begin{equation}\label{OL}
  {\cal L}_{\frac{U(1)_{k^{2}}}{\mathbb{Z}_{k}}}={\cal L}_{U(1)/U(1)}(\widetilde{g},\widetilde{A}).
\end{equation}
The field variables are related to the ones ($g_{U(1)}$, $A_{U(1)}$) in the $U(1)_{k}$ gauged WZW model as follows:
\begin{equation*}
  \widetilde{g}= g^{k}_{U(1)},\quad\quad \widetilde{A}=k\cdot A_{U(1)}.
\end{equation*}
If we denote the variables of $U(k)_{n, n+k}$ to be $g$ and $A$, then we have
\begin{equation*}
  \widetilde{g}=\det g,\quad\quad \widetilde{A}={\rm tr}A.
\end{equation*}
So the partition function of $ SU(n)_{k}$ gauged WZW model can be readily derived:
\begin{eqnarray}\label{PMPF}
   Z(S_{SU(n)_{k}}(\Sigma_{h})) = \frac{1}{k!}\sum^{\infty}_{n_{1},\cdots, n_{k}=-\infty}&&\int\prod^{k}_{a=1}\frac{d\phi_{a}}{2\pi}\prod^{k}_{a,b=1;
   a\neq b}\left(1-e^{2\pi i(\phi_{a}-\phi_{b})}\right)^{1-h}\times \\\nonumber
   && \prod^{k}_{a=1}\delta\left(\frac{n+k}{2\pi}\phi_{a}-n_{a}-\sum_{a}\frac{\phi_{a}}{2\pi}+\frac{k-1}{2}\right) .
\end{eqnarray}
The vacuum equations are then
\begin{equation}\label{VESU}
  \left(g_{a}\right)^{n}=(-1)^{k-1}\frac{e_{k}(g_{a})}{(g_{a})^{k}},\quad\quad {\rm if}\quad a=1,\cdots,k,
\end{equation}
where $e_{k}(g_{a})$ is the elementary symmetric function
\begin{equation}\nonumber
  e_{k}(g_{a})=\prod^{k}_{a=1}g_{a}.
\end{equation}
In order to see the $\mathbb{Z}_{n}$ center symmetry, one needs the gauge-invariant observable $e_{k}(g_{a})$. From Equ'n~(\ref{VESU}), we have
\begin{equation}\label{GVESU}
  \left(e_{k}(g_{a})\right)^{n}=1.
\end{equation}
Indeed, it has $n$ solutions which can be regarded as sites in the bosonic chain integrable system. The chain has a $\mathbb{Z}_{n}$  translation symmetry. Furthermore, Equ'n (\ref{VESU}) is the Bethe ansatz equation of the phase model. The number of solutions of the equations (\ref{VESU}) has been counted in \cite{Korff:2010}:
\begin{equation*}
  \frac{(k+n-1)!}{k!(n-1)!}.
\end{equation*}
A direct evaluation of the sphere partition function gives
\begin{equation}\label{}
   Z(S_{SU(n)_{k}}(S^{2})) =1.
\end{equation}

\subsection{$G/G$ gauged WZW-matter models}\label{GWZWM}
In order to understand the q-deformed Verlinde algebra discussed in \cite{Korff:2011} via the gauge theory, Okuda and Yoshida in \cite{Okuda:2013fea} utilized a  $U(k)/U(k)$ gauged WZW model with adjoint matter; see also \cite[Section 8]{ Gerasimov:2006zt} for an earlier investigation\footnote{We thank S.L. Shatashvili for pointing out this work to us.}. Let us denote the $U(k)$-bundle as $E$. The matter fields introduced in \cite{Okuda:2013fea} are as follows:
\begin{itemize}
    \item Grassmann even and odd sections of the bundle End($E$), denoted $\Phi$ and $\psi$, respectively, 
    \item holomorphic one-form auxiliary fields $\varphi^{(1,0)}\in\Omega^{(1,0)}(\Sigma_{h}, {\rm End}(E))$ (Grassmann even) and $\chi^{(1,0)}\in\Omega^{(1,0)}(\Sigma_{h}, {\rm End}(E))$ (Grassmann odd), 
    \item associated anti-holomorphic fields $\varphi^{(0,1)}$ and $\chi^{(0,1)}$ defined similarly. 
\end{itemize}
The action of the matter fields is
\begin{align*}
  S_{{\rm matter}}&(g,A,\Phi,\Phi^{\dagger},\psi,\psi^{\dagger},\varphi,\chi) \\
  =&-\frac{1}{2\pi} \int_{\Sigma_{h}}{\rm tr}\left(\Phi\Phi^{\dagger}+\psi\psi^{\dagger}-q^{-2}\Phi^{\dagger}g^{-1}\Phi g\right)\\
   &+\frac{1}{4\pi}\int_{\Sigma_{h}}{\rm tr}\{\varphi^{(0,1)}\wedge\left(\partial_{A}\Phi+[X,\Phi]\right)-\chi^{(0,1)}\wedge\left(\partial_{A}\psi+[X,\psi]\right)  \\
   &+ \varphi^{(1,0)}\wedge\left({\bar\partial}_{A}\Phi^{\dagger}-[Y,\Phi^{\dagger}]\right)-\chi^{(1,0)}\wedge\left({\bar\partial}_{A}\psi^{\dagger}-[Y,\psi^{\dagger}]\right) \},
\end{align*}
where
\begin{equation*}
  X=\sum^{\infty}_{l=0}g^{-l}\left(g^{-1}\partial_{A}g\right)g^{l},\quad\quad Y=\sum^{\infty}_{l=0}g^{l}\left({\bar\partial}_{A}gg^{-1}\right)g^{-l},
\end{equation*}
where $g^{l}$ is the $l$-th power of the group variable $g$. The matter action is also BRST-exact, so the partition function of the GWZW-matter model could be computed via the localization technique that has been performed in \cite{Okuda:2013fea}. It is
\begin{equation}\label{GMPF}
  Z_{{\rm GWZWM}}(S_{U(k)_{n}}(\Sigma_{h},q))=\frac{1}{k!}\sum^{\infty}_{m_{1},\cdots,m_{k}=-\infty}\int\prod^{k}_{a=1}\frac{d\phi_{a}}{2\pi}f[\phi_{a},q]\prod^{k}_{a=1}\delta\left(\alpha_{a}\left(\phi\right)-m_{a}\right)
\end{equation}
where $\alpha_{a}\left(\phi\right)$ is defined by
\begin{equation}\label{DVE}
 \alpha_{a}\left(\phi\right)=n\frac{\phi_{a}}{2\pi}-\frac{i}{2\pi}\sum^{k}_{b=1, b\neq a}\log\left(\frac{e^{i\phi_{a}}-q^{-2}e^{i\phi_{b}}}{q^{-2}e^{i\phi_{a}}-e^{i\phi_{b}}}\right)
\end{equation}
 and the measure factor $f[\phi_{a},q]$ is
 \begin{align}\label{MSF}
   f[\phi_{a},q]= &\left|\det\left(\frac{\partial\alpha_{b}\left(\phi\right)}{\partial\phi_{a}}\right)\right|^{h}\cdot\frac{1}{\left(1-q^{-2}\right)^{k(1-h)}}  \\\nonumber
    &\times\left(\prod^{k}_{a,b=1, a\neq b}\frac{e^{i\phi_{a}}-e^{i\phi_{b}}}{e^{i\phi_{a}}-q^{-2}e^{i\phi_{b}}}\right)^{1-h}
 \end{align}
Here, we only present Okuda and Yoshida's results, and interested readers can refer to their paper for the detailed derivation. Similar results were obtained by Gukov and Pei \cite[Section 5.2]{Gukov:2015sna} who studied the partition function of the q-deformed Chern-Simons theory on $\Sigma_{h}\times S^{1}$.  All of the discussions in the above are about the $U(k)_{n}$ GWZW-matter model, however, in order to get the partition function of the $SU(n)_{k}$ GWZW-matter model, we need an extended Seiberg-like duality:
\begin{equation}\label{LRD2}
   SU(n)_{k}+? \leftrightarrow U(k)_{-n,-n}+\textrm{adjoint\ matter }.
\end{equation}
As far as the author knows, there is no such study in the literature yet, and we leave it to the future. Assuming the existence of such duality, we then expect that vacuum structures coincide between the two theories. Thus, Equ'n (\ref{DVE}) is also the vacuum equation of the $SU(n)_{k}$ gauged WZW-matter model that can be rewritten as
\begin{equation}\label{DVE2}
  \left(g_{a}\right)^{n}=\prod^{k}_{b=1, b\neq a}\left(\frac{q^{-2}g_{a}-g_{b}}{g_{a}-q^{-2}g_{b}}\right).
\end{equation}
The central subgroup of $SU(n)$, $\mathbb{Z}_{n}$, is the global symmetry of the $SU(n)_{k}$ gauged WZW-matter model. It can also be read off from vacuum equations in terms of the gauge invariant variable $e_{k}(g_{a})$ since it satisfies $\left(e_{k}(g_{a})\right)^{n}=1$. We will also need the sphere partition function of the $SU(n)$ GWZW-matter theory, which was numerically conjectured by Okuda and Yoshida \cite{Okuda:2013fea} to be
\begin{equation}\label{QDSPF}
  Z_{\textrm{GWZWM}}\left(S_{SU(n)}\left(S^{2},q\right)\right)=\prod^{n}_{a=1}\frac{1}{(1-q^{-2a})}.
\end{equation}
With this conjecture, they reproduced the “deformed Verlinde algebra” constructed by Korff in \cite{Korff:2011} whose construction is motivated by the q-boson model and uses the cylindric generalization of skew Macdonald functions. If we put Korff's with Okuda and Yoshida's work together, it is like constructing gauge theories from the integrable models.

On the other hand, our goal in this paper is to construct the q-boson model from the associated gauge theories guided by the global symmetry, and the only information we will refer to from Equ'n~(\ref{QDSPF}) is that the partition function is not equal to one unless we take the limit where $q$ goes to infinity. This is the limit to the ordinary $SU(n)_{k}$ Verlinde algebra.
The novelty of this work relative to \cite{Bogoliubov:1997soj, Korff:2011} is that, guided by
  the symmetries of the gauge theory, we will propose a new Bethe ansatz in section \ref{NBAQ} to describe the wave function of the integrable model. The partition function (\ref{QDSPF}) is a consequence of the proposed wave function. Furthermore, we will deduce the q-deformed algebra from the proposed wave function. Finally, in order to find the hopping interaction term in the q-boson model from the associated gauge theory, we need to embed  the GWZW-matter theory into a 
  larger theory (the $q$-deformed complex Chern-Simons theory) which we will define in the next section, and 
  that includes the dynamical domain walls.  
  The hopping terms that describe the fluctuations between two ``sites" of the bosonic chain can be mapped to the fluctuations between two vacua of the corresponding gauge theory.
  In the next section, we will give a short discussion about the big theory.

\subsection{The q-deformed complex Chern-Simons matter theory}\label{qdCS}
Before continuing,
let us recall a simpler situation where $q \rightarrow \infty$ (for the $q$ defined in the previous section). In that limit, the supersymmetric gauge theory has only fundamental matter.  Now, a two-dimensional {\cal N}=(2,2) $U(k)$ gauge theory with fundamental matter is the gauged linear sigma model for the Grassmannian $Gr(k,N)$.  As noted by Witten, this GLSM contains in a limit the $U(k)_{N-k, N}$ gauged WZW model, whose Verlinde algebra provides one computation of the quantum cohomology of $Gr(k,N)$.
The vacuum structure of the GLSM depends on the FI parameter $\zeta$:
\begin{itemize}
\item If $\zeta\gg 0$, the semiclassical vacuum configuration can be read off from the D-term
\begin{equation*}
  \{XX^{\dagger}=\zeta\cdot \textbf{I}_{k\times k}\}/U(k).\footnote{\textrm{It was further observed in \cite{Gu:2021beo} that one can label the semiclassical vacuum by using the dynamical theta angle due to the fact that ${\rm tr}F_{01}$ is (pseudo)-scalar in two-dimensional quantum field theory.}}
\end{equation*}
\item If $\zeta\ll 0$, it is $U(k)_{N-k, N}$ gauged WZW model, if we identify the field variables as
\begin{equation*}
  g=c\cdot\sigma,
\end{equation*}
where $\sigma$ is the lowest component field of the superfield strength $\Sigma$, and $c$ is a normalization constant that can be fixed. The gauge field $A$ in the gauged WZW model is the same gauge field $A$ of the nonabelian gauged linear sigma model.
\end{itemize}

We would like to mention two relevant important advances after Witten's paper. The first is the development of mirror symmetry in two-dimensional nonabelian gauged linear sigma models \cite{Gu:2018fpm}.
The second is that at low energies, in two dimensions, the fundamental field $X$ becomes a domain wall between $\sigma$ vacua \cite{Gu:2022ugf}, generalizing the observation in \cite[section IV]{Witten:1978bc} that in two dimensional abelian theories, any state of nonzero charge is a soliton (here, domain wall) in the sense of interpolating between $\sigma$ vacua.  (Classically, $X$ is light, but in a gapped theory it becomes massive, as expected of domain walls.)
It has been explained in \cite{Gu:2022ugf} that these domain walls map to the interaction terms in the Hamiltonians of the integrable system since they both represent the fluctuations between two vacua. Furthermore, we showed that the XX spin chain model emerges from the nonabelian gauged linear sigma model at the intermediate scale \footnote{Let us denote the dynamical scale to be $\Lambda$ which is an RG-invariant quantity, and the gauge coupling to be $e$ which has the mass dimension one in 2d. The intermediate scale is defined to be that $e(\mu>0) \gg \Lambda$, where the physical scale, $\mu$, is not necessary to be vanishing. If it is not at the far infrared, we need to consider the dynamics of the BPS objects. } where the left degrees of freedom are BPS domain walls, which gives a physical explanation of \cite{Korff:2010} where the authors have constructed the quantum cohomology of Grassmannian from the XX model.

However, the vacuum structure of $SU(n)_{k}$ gauged WZW-(matter) models suggests that they cannot be embedded into the usual 2d gauged linear sigma model. 
Instead, they are embedded in a different theory, namely the $\beta$-deformed complex Chern-Simons theory \cite[section 5]{Gukov:2015sna} (motivated by \cite{Kapustin:2013hpk}) specialized to the three-manifold $\Sigma_h \times S^1$.  The $\beta$-deformed complex Chern-Simons theory is given by
\begin{align*}
 3d &\textrm{ ${\cal N}$=2 $U(k)$  Chern-Simons-matter theory at level $\frac{n}{2}$ }\\
   &+\textrm{ n chiral fundamental multiplets X} \\
   &+ \textrm{1 massive chiral multiplet $\Phi$ in the adjoint representation with mass $\beta$}.
\end{align*}
We refer to this theory, the $\beta$-deformed Chern-Simons theory on $\Sigma_h \times S^1$, as $q$-deformed complex Chern-Simons theory, where $q=e^{\pi R\beta}$, for $R$ the radius of the $S^{1}$.
The phases are given as follows:
\begin{itemize}
\item If $\zeta\gg 0$, the D-flatness condition says the Higgs branch is
\begin{equation*}
  \{XX^{\dagger}+[\varphi,\varphi^{\dagger}]=\zeta\cdot\textbf{I}_{k\times k}\}/U(k).
\end{equation*}
\item  If $\zeta\ll 0$, after integrating out X fields, the low-energy effective theory is ${\cal N}$=2 $U(k)_{n}$ Chern-Simons-matter theory with an adjoint chiral superfield $\Phi$ of real mass $\beta$. 
\end{itemize}
Assuming the R-charge of the adjoint matter vanishes, if we further integrate out the massive fermion, 
it reduces to the $U(k)_{n}$ GWZW-matter model
of the previous section, and the $q = e^{\pi R \beta}$ also coincides with the $q$ there. 
Finally, we want to point out that 
\cite{Gukov:2015sna}
also discussed the situation where the adjoint matter has an R-charge 2, and they found the \textit{equivariant Verlinde algebra} in this setup. A different R-charge of adjoint matter will not change the vacuum equations, but it affects the sphere partition function of the GWZW-matter model. Interested readers can read more about the equivariant Verlinde algebra in \cite{Gukov:2015sna,Kanno:2018qbn}.

In the next section, we will get a q-boson model from the gauge theory reviewed in this section by proposing a new Bethe ansatz of wave function inspired by symmetries of the associated gauge theory. Our approach might also shed new light on the connection between the integrable system and the equivariant Verlinde algebra.

\section{A new Bethe ansatz for q-boson models}\label{NBAQ}
In previous studies \cite{Bogoliubov:1997soj, Korff:2011}, the wave functions (or Bethe eigenvectors) were built from the entries of the monodromy matrix $T(g)$ that is used to describe the quantum integrability of the q-boson model:
\begin{equation*}
  T(g)=\left(
                           \begin{array}{cc}
                             A(g) & B(g) \\
                             C(g) & D(g) \\
                           \end{array}
                         \right).
\end{equation*}
Then eigenvectors were proposed to be
\begin{equation}   \label{eq:wavefn}
  |\psi_{n}\left(g_{1},\cdots,g_{n}\right)\rangle=\prod^{n}_{a=1}B\left(g_{a}\right)|0\rangle.
\end{equation}
where $|0\rangle$ is the normalized ground states on the n-sites lattice and the parameter $g_{a}$ satisfy the Bethe (vacuum) equations (\ref{DVE2}). This wave function has a factorized structure. In this section, we will propose a different Bethe ansatz for the wave function inspired by the symmetries of $SU(n)_{k}$ GWZW-matter theory. First, let us recall how we use the symmetries of quantum gauge theory to get the wave function of Heisenberg spin chain as was studied in \cite{Gu:2022ugf}. For the $\textrm{XXX}_{s}$ spin chain, the vacuum equations are
\begin{equation*}
  \left(\frac{g_{a}+is}{g_{a}-is}\right)^{N}=\prod^{k}_{b\neq a}\frac{g_{a}-g_{b}+i}{g_{a}-g_{b}-i},\quad\quad 1\leq a\leq k.
\end{equation*}
We can write out the vacuum equations in terms of the gauge invariant variable
\begin{equation*}
  \left(e_{k}\left(\frac{g_{a}+is}{g_{a}-is}\right)\right)^{N}=1
\end{equation*}
which manifest the $\mathbb{Z}_{N}$ global symmetry of the UV gauge theory. By requiring that the wave function~(\ref{eq:wavefn}) be neutral under that ${\mathbb Z}_N$ and also the
residual gauge symmetry $S_k$ (the Weyl group of $U(k)$), and also requiring that the wave function be an eigenvalue of the Hamiltonians, we can recover known results for the $\textrm{XXX}_{s}$ spin chain, which we outline below. \\

\noindent \underline{$\textrm{XXX}_{\frac{1}{2}}$ model}. The wave function is well-known:
\begin{equation*}
    \psi\left(p_{1},\cdots, p_{k}\right)=A_{k}\sum_{1\leq j_{1}<\cdots <j_{k}\leq N}\sum_{{\cal W}\in S_{k}}S_{{\cal W} \left(a\right)}e^{\sum^{k}_{a=1}ip_{{\cal W} \left(a\right)}j_{a}}|j_{1}<\cdots< j_{k}\rangle_{k},
\end{equation*}
where 
\begin{equation*}
  e^{ip_{a}}=\frac{g_{a}+\frac{i}{2}}{g_{a}-\frac{i}{2}},
  \: \: \:
  S_{ab}=\frac{g_{a}-g_{b}+i}{g_{a}-g_{b}-i},
\end{equation*}
and $|j_{1}<\cdots< j_{k}\rangle_{k}$ denotes a configuration that has $k$ spin up sites, and spin down for the rest $N-k$ sites.

The wave function obeys the periodic boundary condition
\begin{equation}\label{PBC}
  |j_{2}<\cdots<j_{k}<j_{1}+N>= |j_{1}<j_{2}<\cdots<j_{k}>.
\end{equation}
Let us list several concrete examples: \\

\begin{itemize}
    \item  If $k=2$, we have
\begin{equation*}
  A_{2}\sum_{1\leq j_{1}<j_{2}\leq N}\left(e^{i\left(p_{1}j_{1}+p_{2}j_{2}\right)}+S_{21}e^{i\left(p_{2}j_{1}+p_{1}j_{2}\right)}\right)|j_{1}<j_{2}\rangle.
\end{equation*}
\item If $k=3$, we have
\begin{align*}
   |p_{1},p_{2},p_{3}\rangle=&|p_{1}<p_{2}<p_{3}\rangle+S_{21}S_{31}S_{32}|p_{3}<p_{2}<p_{1}\rangle \\
   &+S_{21}|p_{2}<p_{1}<p_{3}\rangle+S_{31}S_{32}|p_{3}<p_{1}<p_{2}\rangle \\
   &+S_{32}|p_{1}<p_{3}<p_{2}\rangle+S_{21}S_{31}|p_{2}<p_{3}<p_{1}\rangle,
\end{align*}
where the notation $|p_{a}<p_{b}<p_{c}\rangle$ represents
\begin{equation*}
  |p_{a}<p_{b}<p_{c}\rangle=A_{3}\sum_{1\leq j_{1}<j_{2}<j_{3}\leq N}\e^{i\left(p_{a}j_{1}+p_{b}j_{2}+p_{c}j_{3}\right)}|j_{1}<j_{2}<j_{3}\rangle.
\end{equation*}
\end{itemize}

\noindent \underline{XXX model with higher spin}. Instead of giving a general proposal for the wave function, we work with a minor modification of the case $s=1$:\\

\begin{itemize}
    
\item  $k=0$, i.e, vacuum:
\begin{equation*}
  |0\rangle=|0\cdots0\rangle.
\end{equation*}
\item  $k=1$, i.e, one-magnon states:
\begin{equation*}
  |p\rangle=\sum_{j}e^{ipj}|\cdots \stackrel{j}{1}\cdots\rangle,
\end{equation*}
where we use the notation $|\cdots \stackrel{j}{n}\cdots\rangle$ to indicate $n$ creation operators at site $j$ (and no creation operators at other sites).

\item  $k=2$, i.e, two-magnon states:
\begin{align*}
  |p_{1}< p_{2}\rangle= &\sum_{j_{1}<j_{2}}e^{i(p_{1}j_{1}+p_{2}j_{2})}|\cdots \stackrel{j_{1}}{1}\cdots\stackrel{j_{2}}{1}\cdots\rangle,  \\
   |p; 2\rangle=& \sum_{j}e^{ipj}|\cdots \stackrel{j}{2}\cdots\rangle.
\end{align*}
\end{itemize}
Unlike the $s=1/2$ case, we could have a state in which two magnons are located on the same site when $s > 1/2$.
In order to be an eigenstate of the Hamiltonian of the XXX spin chain, the scattering ansatz needs to be modified by including the extra states as follows:
\begin{equation*}
  |p_{1},p_{2}\rangle= |p_{1}<p_{2}\rangle+S_{21} |p_{2}<p_{1}\rangle+C|p_{1}+p_{2},2\rangle.
\end{equation*}

The coefficient $S_{21}$ is the usual scattering factor.
The term $C$ describes collisions of two magnons on a single site, 
which is referred to as a contact term and crucial for the solution in the integrable system.

So what can we learn from the above higher spin Heisenberg model for constructing the wave function of the q-boson model? We first expect that the corresponding integrable system would be the bosonic chain. It is because the matter and group-valued fields in GWZW-(matter) model are all bosonic fields. As a bosonic system, two creation operators can be put into the same site. Unlike the higher spin chain case in the XXX model, this, in fact, affects the IR physics of the bosonic system, as this contact configuration is also a vacuum state in the $SU(n)_{k}$ gauged WZW-matter model. Nevertheless, the properties of the wave function of the $k=2$ $\textrm{XXX}_{1}$ model give us a hint to propose the wave function of the q-boson model.

Eventually, we arrive at a new Bethe ansatz for the Bethe eigenstates of the q-boson model associated with the $SU(n)_{k}$ gauged WZW-matter model as follows:
\begin{eqnarray}\label{qbwf}
   \psi\left(p_{1},\cdots, p_{k}\right) &=& \sum_{1\leq j_{1}\leq\cdots \leq j_{k}\leq n}a(j_{1},\cdots,j_{k})|j_{1}\leq\cdots\leq j_{k}\rangle_{k} \\
   &=& A_{k}\sum_{1\leq j_{1}\leq\cdots \leq j_{k}\leq n}\sum_{{\cal W}\in S_{k}}S_{{\cal W} \left(a\right)}e^{\sum^{k}_{a=1}ip_{{\cal W} \left(a\right)}j_{a}}|j_{1}\leq\cdots\leq j_{k}\rangle_{k},
\end{eqnarray}
where 
\begin{equation*}
  e^{ip_{a}}=g_{a}, \quad  \quad S_{ab}=\frac{q^{-2}g_{a}-g_{b}}{g_{a}-q^{-2}g_{b}},
\end{equation*}
 $S_{k}$ is the Weyl group of $U(k)$, and $A_{k}$ is an overall normalization constant. The ket 
 \begin{equation*}
     |j_{1}\leq\cdots\leq j_{k}\rangle_{k}
 \end{equation*}
 denotes a particular ground state of the gauge theory. However, as mentioned before, the bosonic chain will appear at the intermediate scale of the gauge theory, where only ground states and domain walls are left. So, the state at this scale should be neutral under all symmetries, indicating no spontaneous symmetry breaking here. There could be several possible representations of this state once one can show that it is an eigenstate of the Hamiltonian by using the Bethe (or vacuum) equations. This \textit{criterion}  will be the consistency condition for our new observation. Indeed, the new proposed Bethe ansatz is $\mathbb{Z}_{n}$ neutral and gauge-invariant under the Weyl symmetry $S_{k}$. Before verifying this proposal, let us comment on the scattering factor we choose for the q-boson model first. When we take the identical momenta limit of $S_{ab}$ , we have
\begin{equation*}
  S(p,p)=-1.
\end{equation*}
This indicates that the particles obey Fermi statistics, which naively appears to contradict our claim that the integrable model is a bosonic system. In fact, the first attempt of the scattering factor by the author was
\begin{equation}\nonumber
  \widetilde{S}_{ab}=-\frac{q^{-2}g_{a}-g_{b}}{g_{a}-q^{-2}g_{b}}.
\end{equation}
However, it does not satisfy the criteria mentioned before. Our choice of scattering factor S suggests fermionic excitations in our bosonic system. We leave the explanation of this for the future.  
We observe that this is analogous to fermionization in two dimensions.

In the next section, we will verify the proposal for the integrable system, and we will find a q-deformed algebra emerging from the consistency conditions of the integrability.

\section{The q-deformed algebra}\label{QDA}
As we observed in section \ref{PGT}, the $SU(n)_{k}$ GWZW-matter model has a global $\mathbb{Z}_{n}$ symmetry that can be seen as a phase rotation between solutions to the vacuum equation
\begin{equation*}
  \left(e_{k}(g_{a})\right)^{n}=1.
\end{equation*}
This equation is satisfied for any $k$ so that one can introduce a unit-less circle variable $\tau$ satisfying $\tau^{n}=1$, and its $n$ solutions will be regarded as the $n$ sites of the q-deformed model. Following the construction in \cite{Gu:2022ugf}, we assign a pair of creation and annihilation operators on each site for the integrable system, and they are both charged under the $\mathbb{Z}_{n}$ symmetry. Let us denote them as
\begin{equation*}
  Q_{i},\quad Q^{\dagger}_{i}.
\end{equation*}
These operators in the associated integrable system could be used to construct the operators of the q-deformed complex Chern-Smions on $\Sigma\times S^1$. However, they are not the operators in gauge theory.

Since one may naturally expect that
\begin{equation*}
  [Q_{i}, Q^{\dagger}_{i}]\sim 1,
\end{equation*}
 $Q_{i}$ and $Q^{\dagger}_{i}$ should have opposite $\mathbb{Z}_{n}$-charges with the same magnitude. As observed in \cite{Gu:2022ugf},  the Hamiltonian $H_l$ of the q-boson model 
 generates
 shifts of the $i$-th site to the $(i+l)$-th site. The simplest basis of the Hamiltonians would be
\begin{equation}\label{HDQB}
  H_{l}=\sum^{n}_{i=1}\left(Q_{i+l}Q^{\dagger}_{i}+Q_{i}Q^{\dagger}_{i+l}\right),\quad\quad  l=1,\cdots,n.
\end{equation}
Each term $Q_{i+l}Q^{\dagger}_{i}$ is an operator in the UV gauge theories reviewed before, which represents a domain wall. However, since those Hamiltonians commute with each other, their linear combinations could also be regarded as ``Hamiltonians" of the system. The detailed information of the Hamiltonians certainly depends on the commutators of $Q$ and $Q^{\dagger}$ and the interaction between bosons. However, we do not know the commutators yet. So our strategy in this paper is to first choose a ``gauge" of the fundamental Hamiltonian by requiring that it is just $H_{1}$ in (\ref{HDQB}), and then derive the local commutators by the consistency condition: the wave function (\ref{qbwf}) shall be an eigenstate of the Hamiltonian
 \begin{equation*}
  H=\sum^{n}_{i=1}\left(Q_{i+1}Q^{\dagger}_{i}+Q_{i}Q^{\dagger}_{i+1}\right).
\end{equation*}

Let us investigate the commutators case by case and focus solely on the holomorphic part of the fundamental Hamiltonian $Q_{i+1}Q^{\dagger}_{i}$ that moves each boson to its left adjacent site.

\begin{itemize}
    \item  For the case $k=2$, the operator $\sum_{i}Q_{i+1}Q^{\dagger}_{i}$ changes the local states as follows:
\begin{align*}
    \mid\cdots\stackrel{j_{1}+1}{1}\cdots\stackrel{j_{2}}{1}\cdots\rangle,\ \mid\cdots\stackrel{j_{1}}{1}\cdots\stackrel{j_{2}+1}{1}\cdots\rangle &\mapsto \mid\cdots\stackrel{j_{1}}{1}\cdots\stackrel{j_{2}}{1}\cdots\rangle  \quad {\rm for}\quad j_{1}<j_{2}\\
   \mid\cdots\stackrel{j_{1}}{1}\stackrel{j_{1}+1}{1}\cdots\rangle&\mapsto \mid\cdots\stackrel{j_{1}}{2}\cdots\rangle,
\end{align*}
\end{itemize}
where we use the notation $\mid\cdots\stackrel{j_{1}}{1}\cdots\stackrel{j_{2}}{1}\cdots\rangle$ here instead of  $\mid j_{1}<j_{2}\cdots\rangle$ to make the action of $\sum_{i}Q_{i+1}Q^{\dagger}_{i}$ more transparent. One may naively impose the following local algebra:
\begin{equation*}\label{}
  Q\cdot|m\rangle=|m-1\rangle,\quad\quad Q^{\dagger}\cdot|m\rangle=|m+1\rangle,
\end{equation*}
where $|m\rangle=0$ if $m<0$.  From Equ'n~(\ref{qbwf}), we see that the wave function of the case $k=2$ is 
\begin{equation*}
 A_2 \sum_{1\leq j_1 \leq j_2 \leq N}\left (e^{i(p_1j_1+p_2j_2)}+S_{21}e^{i(p_2j_1+p_1j_2)}\right)\mid j_1 \leq j_2\rangle.
\end{equation*}
Then we have the following identities:
\begin{align*}
  a(j_{1}+1,j_{2})+a(j_{1},j_{2}+1) & =f_{1}(g_{1},g_{2})\cdot a(j_{1},j_{2}) \\
   a(j_{1},j_{1}+1) & =f_{2}(g_{1},g_{2})\cdot a(j_{1},j_{1}).
\end{align*}
The new Bethe ansatz could be an eigenstate of the Hamiltonian if $f_{1}(g_{1},g_{2})=f_{2}(g_{1},g_{2})$. However, by direct computation, we have 
 \begin{eqnarray*} a(j_1+1,j_2)&+&a(j_1,j_2+1) \\\nonumber
 &=&(g_1+g_2)\left(e^{i(p_1j_1+p_2j_2)}+S_{21} e^{i(p_2j_1+p_1j_2)}\right)\\\nonumber
  &=&(g_1+g_2)a(j_1,j_2),
\end{eqnarray*}
and 
 \begin{eqnarray*}  a(j_1,j_1+1) &=& g_2e^{i(p_1j_1+p_2j_1)}+g_1S_{21}e^{i(p_2j_1+p_1j_1)}\\\nonumber
&=& (g_2+g_1S_{21})e^{i(p_1j_1+p_2j_1)} \\\nonumber
&=& \frac{g_1+g_2}{1+q^{-2}} (1+S_{21})e^{i(p_1j_1+p_2j_1)}\\\nonumber
&=& \frac{g_1+g_2}{1+q^{-2}}a(j_1,j_2).
\end{eqnarray*}

This tells \begin{equation*}
  f_{1}(g_{1},g_{2})=g_{1}+g_{2},\quad\quad f_{2}(g_{1},g_{2})=\frac{g_{1}+g_{2}}{1+q^{-2}}.
\end{equation*}
So the previous proposed local algebra can not be correct. To be consistent, we require a modified local algebra
\begin{equation}\label{}
   Q\cdot|m\rangle=l_{m}|m-1\rangle,\quad\quad Q^{\dagger}\cdot|m\rangle=n_{m}|m+1\rangle.
\end{equation}
Then, the identities will be modified to
\begin{align*}
 l_{1}n_{0}\left( a(j_{1}+1,j_{2})+a(j_{1},j_{2}+1)\right) & =\left(g_{1}+g_{2}\right)\cdot a(j_{1},j_{2}) \\
   l_{1}n_{1}a(j_{1},j_{1}+1) & =\left(g_{1}+g_{2}\right)\cdot a(j_{1},j_{1}).
\end{align*}
So we should have
\begin{equation*}
  l_{1}n_{0}=1, \quad\quad l_{1}n_{1}=1+q^{-2},
\end{equation*}
which implies
\begin{equation*}
  \frac{n_{1}}{n_{0}}=1+q^{-2}.
\end{equation*}
A plausible subalgebra would be modified to
\begin{equation}\label{LV2}
  Q\cdot|1\rangle=|0\rangle,\quad\quad Q^{\dagger}\cdot|0\rangle=|1\rangle,\quad\quad Q^{\dagger}\cdot|1\rangle=(1+q^{-2})|2\rangle.
\end{equation}

\begin{itemize}
    \item  Now, let us move to the $k=3$ case. The operator $\sum_{i}Q_{i+1}Q^{\dagger}_{i}$ acts on the local states as follows:
\begin{align*}
    \mid\cdots\stackrel{j_{1}+1}{1}\cdots\stackrel{j_{2}}{1}\cdots\stackrel{j_{3}}{1}\cdots\rangle,\ &\mid\cdots\stackrel{j_{1}}{1}\cdots\stackrel{j_{2}+1}{1}\cdots\stackrel{j_{3}}{1}\cdots\rangle,\ \mid\cdots\stackrel{j_{1}}{1}\cdots\stackrel{j_{2}}{1}\cdots\stackrel{j_{3}+1}{1}\cdots\rangle  \\ \mapsto \mid\cdots\stackrel{j_{1}}{1}&\cdots\stackrel{j_{2}}{1} \cdots\stackrel{j_{3}}{1}\cdots\rangle  \quad {\rm for}\quad j_{1}<j_{2}<j_{3},
 \end{align*}
 \begin{align*}
\mid\cdots\stackrel{j_{1}+1}{1}\cdots\stackrel{j_{ 2}}{2}\cdots\rangle,\ \mid\cdots\stackrel{j_{1}}{1}\cdots\stackrel{j_{ 2}}{1}\stackrel{j_{ 2}+1}{1}\cdots\rangle& \mapsto \mid\cdots\stackrel{j_{1}}{1}\cdots\stackrel{j_{2}}{2}\cdots\rangle  \quad {\rm for}\quad j_{1}<j_{2},\\
      \mid\cdots\stackrel{j_{1}}{2}\stackrel{j_{ 1}+1}{1}\cdots\rangle &\mapsto \mid\cdots\stackrel{j_{1}}{3}\cdots\rangle.
   \end{align*}
   \end{itemize}
If we write those relations in terms of coefficients, we would have
\begin{align*}
  a(j_{1}+1,j_{2},j_{3})+ a(j_{1},j_{2}+1,j_{3})+ a(j_{1},j_{2},j_{3}+1) &= f_{1}(g_{1},g_{2},g_{3})a(j_{1},j_{2},j_{3}),  \\
   a(j_{1}+1,j_{2},j_{2})+ \left(1+q^{-2}\right)a(j_{1},j_{2},j_{2}+1) &= f_{2}(g_{1},g_{2},g_{3})a(j_{1},j_{2},j_{2}),\\
    n_{2}a(j_{1},j_{1},j_{1}+1) &= f_{3}(g_{1},g_{2},g_{3})a(j_{1},j_{1},j_{1}).
\end{align*}
A direct computation gives
\begin{equation*}
  f_{1}(g_{1},g_{2},g_{3})=f_{2}(g_{1},g_{2},g_{3})=f_{3}(g_{1},g_{2},g_{3})=g_{1}+g_{2}+g_{3},\quad{\rm and}\quad l_{2}=1, \quad n_{2}=1+q^{-2}+q^{-4}.
\end{equation*}
This strongly suggests that there is a q-deformed algebra present.

\begin{itemize}
\item  For general $k$, using the identities discussed in Appendix \ref{IDAP}, we arrive at the following:
\begin{equation}\label{QDAI}
   Q\cdot|m\rangle=|m-1\rangle,\quad\quad Q^{\dagger}\cdot|m\rangle=[m+1]|m+1\rangle,
\end{equation}
\end{itemize}
where $[m]$ is
\begin{equation*}
  [m]=\frac{1-q^{-2m}}{1-q^{-2}}.
\end{equation*}

Now, if we define the number operators $N_{i}$ with the following algebra:
\begin{equation}\label{NOA}
  [N_{i}, Q^{\dagger}_{j}]=Q^{\dagger}_{j}\delta_{ij},\quad\quad [N_{i}, Q_{j}]=-Q_{j}\delta_{ij}.
\end{equation}
We then have $N\cdot|m\rangle=m|m\rangle.$

Eventually, a full q-deformed boson algebra emerges
\begin{align}\label{qdba}
  q^{-N}q^{N}=q^{N}q^{-N}=1, \quad q^{-N}Q&=Qq^{-(N-1)},\quad\quad q^{-N}Q^{\dagger}=Q^{\dagger}q^{-(N+1)},\\
  [Q_{i},Q_{j}^{\dagger}]=q^{-2N_{i}}\delta_{ij},\quad&\quad  Q_{i}Q_{j}^{\dagger}-q^{-2}Q_{j}^{\dagger}Q_{i}=\delta_{ij}.
\end{align}
This algebra is also related to $SU_{q}(2)$ algebra \cite{Kulish:1981}.

The Fock space can be then easily constructed as follows:
\begin{equation}\label{LFS}
  |m_{1},\cdots, m_{n}\rangle=\prod^{n}_{j=1}\left([m_{j}]!\right)^{-1}\left(Q^{\dagger}_{j}\right)^{m_{j}}|0\rangle.
\end{equation}

Compared to the algebras in \cite[Equ'n(3.1) to Equ'n(3.6)]{Korff:2011}, the only difference is the following
\begin{equation*}
  Q^{\dagger}\mapsto \frac{\beta^{\dagger}}{1-q^{-2}}.
\end{equation*}
Comparing with \cite{Bogoliubov:1997soj}, the algebras are exactly the same. However, the Fock spaces are different. The state space defined in \cite{Bogoliubov:1997soj} is
 \begin{equation*}
  \widetilde{|m\rangle}=\prod^{n}_{j=1}\left([m_{j}]!\right)^{-\frac{1}{2}}\left(Q^{\dagger}_{j}\right)^{m_{j}}|0\rangle.
\end{equation*}
Its norm is one that has an overall normalization difference from our proposed state. We will argue that our choice is more ``natural" from the perspective of quantum field theory.  Recall the partition function (\ref{QDSPF}),
\begin{equation*}
  Z_{\textrm{GWZWM}}\left(S_{SU(n)}\left(S^{2},q\right)\right)=\prod^{n}_{a=1}\frac{1}{(1-q^{-2a})}
  \end{equation*}
  is related to the norm of the state space, and more details about the connection can be found in \cite{Korff:2011, Okuda:2013fea}. Since it is not equal to one unless $q$ is infinity, we expect that the Fock state emerging from the gauge theory should not have a unit norm.

  Finally, the eigenvalue of the Hamiltonian is
  \begin{equation*}
    H\cdot|k\rangle=\left(g_{1}+\cdots+g_{k}+{\bar g}_{1}+\cdots+{\bar g}_{k}\right)|k\rangle.
  \end{equation*}
  A similar computation applies to higher Hamiltonians (need to be constructed) as well, and their eigenvalues would be the complete homogeneous symmetric polynomials
  \begin{equation*}
    h_{a}(g_{1},\cdots, g_{k})+h_{a}\left({\bar g}_{1},\cdots, {\bar g}_{k}\right) \quad\quad a=2,\cdots, k.
  \end{equation*}


  \begin{center}
\section*{Acknowledgement}
\end{center}

We thank Sergei Gukov, Florian Loebbert, Du Pei, and Xingyang Yu for discussions. We would also like to thank the organizers of GLSM@30: Mykola Dedushenko, Heeyeon Kim, Johanna Knapp, Ilarion Melnikov, and Eric Sharpe for inviting the author to be a speaker and asking the author to contribute this article to this conference. Finally, we thank the Simons Center for Geometry and Physics for supporting the excellent workshop
GLSM@30, which made possible many stimulating discussions.
\\

 \appendix
  \section{Identities}\label{IDAP}
For a general $k$, the holomorphic part of the fundamental Hamiltonian gives the following map between the local Fock spaces, for example,
\begin{align*}
  |\cdots\stackrel{j}{k-1}\stackrel{j+1}{1}\cdots\rangle & \mapsto |\cdots\stackrel{j}{k}\cdots\rangle, \\
  |\cdots\stackrel{j}{k}\cdots\rangle, \ |\cdots \stackrel{j-1}{1}\stackrel{j}{k-2}\stackrel{j+1}{1}\cdots\rangle & \mapsto |\cdots\stackrel{j-1}{1}\stackrel{j}{k-1}\cdots\rangle.
\end{align*}
 The associated identities induced from the above we will prove are
 \begin{align}\label{IDEE}
 [k]a(\stackrel{k-1}{\overbrace{j,\cdots,j}},j+1)&= (g_{1}+\cdots+g_{k})a(j,\cdots,j),\\
 a(j,\cdots,j)+[k-1]a(j-1,\stackrel{k-2}{\overbrace{j,\cdots,j}},j+1)  &= (g_{1}+\cdots+g_{k})a(j-1,\stackrel{k-1}{\overbrace{j,\cdots,j}}).
 \end{align}
These two identities teach us that
\begin{equation*}
  l_{k}=1\quad {\rm and} \quad n_{k}=[k+1],
\end{equation*}
which can be proved by induction. To see this, we first assume that
\begin{equation*}
a(\stackrel{k-1}{\overbrace{j,\cdots,j}})=[k-1]!\prod^{k-1}_{a=1}\left(g_{a}\right)^{j}\frac{\prod_{1\leq a<b\leq k-1}\left({g_{a}-g_{b}}\right)}{\prod_{1\leq a<b\leq k-1}\left({q^{-2}g_{a}-g_{b}}\right)}.
\end{equation*}
Based on the property of the $S_{k}$ group \cite{WIKI}, we have
\begin{align*}
   a(\stackrel{k}{\overbrace{j,\cdots,j}})=&\left(g_{1}\right)^{j}a_{1}(\stackrel{k-1}{\overbrace{j,\cdots,j}})+\left(g_{2}\right)^{j}S_{21}a_{2}(\stackrel{k-1}{\overbrace{j,\cdots,j}})&  \\
   &+\cdots+\left(g_{k}\right)^{j}\prod^{k-1}_{b=1}S_{kb}a_{k}(\stackrel{k-1}{\overbrace{j,\cdots,j}}),
\end{align*}
where the subscript in $a_{l}(\stackrel{k-1}{\overbrace{j,\cdots,j}})$ means that its variables are the set:
\begin{equation*}
\{g_{1},\cdots,g_{l-1},g_{l+1},\cdots,g_{k}\}.
\end{equation*}
A tedious but straightforward computation tells
 \begin{equation*}
a(\stackrel{k}{\overbrace{j,\cdots,j}})=[k]!\prod^{k}_{a=1}\left(g_{a}\right)^{j}\frac{\prod_{1\leq a<b\leq k}\left({g_{a}-g_{b}}\right)}{\prod_{1\leq a<b\leq k}\left({q^{-2}g_{a}-g_{b}}\right)}.
\end{equation*}
In the above calculation, one may need the following identity:
\begin{equation}\label{IDE}
  \sum^{k}_{l=1}\Delta^{k-1}_{l}\prod_{1\leq b<l}\left(g_{b}-q^{-2}g_{l}\right)\prod_{l<c\leq k}\left(q^{-2}g_{l}-g_{c}\right)=[k]\Delta^{k},
\end{equation}
where $\Delta^{k}$ is a Vandermonde determinant
\begin{equation*}
  \Delta^{k}=\prod_{1\leq a<b\leq k}\left({g_{a}-g_{b}}\right).
\end{equation*}
The subscript $l$ in $\Delta^{k-1}_{l}$ has the same meaning as the one in $a_{l}(\stackrel{k-1}{\overbrace{j,\cdots,j}})$. The author of this paper showed the identity (\ref{IDE}) by expanding the left-hand side and reorganizing it.

A similar calculation suggests
\begin{equation*}
  a(\stackrel{k-1}{\overbrace{j,\cdots,j}},j+1)=[k-1]!\prod^{k}_{a=1}\left(g_{a}\right)^{j}\left(g_{1}+\cdots+g_{k}\right)\frac{\prod_{1\leq a<b\leq k}\left({g_{a}-g_{b}}\right)}{\prod_{1\leq a<b\leq k}\left({q^{-2}g_{a}-g_{b}}\right)}.
\end{equation*}

Thus, we have proved the first identity appearing in Equ'n (\ref{IDEE}). The second identity can be verified in the same manner. In order to show that our new Bethe ansatz is the correct one, we have to study many more identities than the two in Equ'n (\ref{IDEE}), and the author did not find a universal way to present them. Since their calculations are rather complicated and straightforward, we will not list them here. Finally, we end this appendix by pointing out that one can also check our Bethe ansatz by a numerical simulation.

\end{document}